\documentclass[final,5p,times,twocolumn]{elsarticle}

\usepackage{amsmath}
\usepackage{amssymb}

\biboptions{numbers,sort&compress}

\newcommand{\bmath}[1]{\mbox{\boldmath{$#1$}}}

\begin{document}

\begin{frontmatter}

\title{Instability of sphaleron black holes in asymptotically anti-de Sitter space-time}

\author{Elizabeth Winstanley}
\ead{E.Winstanley@sheffield.ac.uk}
\address{Consortium for Fundamental Physics, School of Mathematics and Statistics, University of Sheffield,\\
Hicks Building, Hounsfield Road, Sheffield S3 7RH, United Kingdom}
\address{Department of Physics and Astronomy, University of Canterbury, Private Bag 4800, Christchurch 8140, New Zealand}

\date{\today}

\begin{abstract}
We prove that sphaleron black holes in ${\mathfrak {su}}(2)$ Einstein-Yang-Mills-Higgs theory with a Higgs doublet in four-dimensional, asymptotically anti-de Sitter space-time are unstable.
\end{abstract}

\begin{keyword}
Black hole, Einstein-Yang-Mills-Higgs theory, anti-de Sitter space-time
\PACS 04.40.-b \sep 04.70.bw
\end{keyword}

\end{frontmatter}

\section{Introduction}
\label{sec:intro}

There is now a zoo of soliton and hairy black hole solutions of Einstein-Yang-Mills (EYM) theory and its variants in both asymptotically flat and asymptotically anti-de Sitter (adS) space-times (see \cite{Volkov:1998cc,Winstanley:2008ac,Winstanley:2015loa,Volkov:2016ehx} for some reviews).
In pure EYM in four space-time dimensions with gauge group ${\mathfrak {su}}(2)$, all nontrivial, asymptotically flat, soliton \cite{Bartnik:1988am} and black hole solutions \cite{Volkov:1989fi,Volkov:1990sva,Bizon:1990sr,Kuenzle:1990is} are unstable \cite{Straumann:1989tf,Straumann:1990as,Lavrelashvili:1994rp,Volkov:1995np,Hod:2008ir}.
With an appropriate choice of gauge, linear, spherically symmetric, perturbations of the metric and gauge field decouple into two sectors, with different properties under a parity transformation: an even-parity (or gravitational) sector and an odd-parity (or sphaleronic) sector \cite{Lavrelashvili:1994rp}.
In the odd-parity sector, instability of the solitons \cite{Galtsov:1991du} and black holes \cite{Volkov:1994dq} can be proven by an elegant method using a variational technique; this does not require knowledge of the details of the equilibrium solutions, just their global behaviour and the boundary conditions on the fields.
The instability in the odd-parity sector is similar to that of the flat-space electroweak sphaleron in ${\mathfrak {su}}(2)$ Yang-Mills-Higgs (YMH) theory \cite{Manton:1983nd,Klinkhamer:1984di,Burzlaff:1983jb,Yaffe:1989ms,Brihaye:1990mw,Brihaye:1992ik} (hence the moniker ``sphaleronic sector''), leading to a sphaleron interpretation of the soliton solutions \cite{Galtsov:1991du}.

Given this analogy between the flat-space YMH sphaleron and solutions of EYM theory, it is interesting to study gravitating solitons and black holes in Einstein-Yang-Mills-Higgs (EYMH) theory.
In four-dimensional asymptotically flat space-time, with gauge group ${\mathfrak {su}}(2)$, sphaleron-like solutions have a doublet-Higgs field in the fundamental representation of the gauge group.
Static, spherically symmetric, soliton and black hole equilibrium solutions of EYMH were studied numerically in \cite{Greene:1992fw}. There are two families of solutions, both of which share features with the pure EYM solutions (to which they reduce when the Higgs coupling is turned off).
Like the EYM solutions, both the solitons \cite{Boschung:1994tc} and black holes \cite{Winstanley:1995iq,Mavromatos:1995kc} in EYMH have a sphaleron-like instability in the odd-parity sector of linear, spherically symmetric perturbations.

It is well-known that the properties of EYM solitons and black holes in asymptotically adS space-time are radically different from those of the corresponding solutions in asymptotically flat space-time.
In particular, for ${\mathfrak {su}}(2)$ gauge group, there exist nontrivial EYM solitons and black holes which are stable under linear, spherically symmetric perturbations in both the odd-parity and even-parity sectors \cite{Winstanley:1998sn,Bjoraker:1999yd,Bjoraker:2000qd}.

The following question then arises: does this existence of stable pure EYM solutions in adS extend to EYMH solitons and black holes?
Numerical solutions of the ${\mathfrak {su}}(2)$ EYMH equations, with a doublet Higgs field, in four-dimensional asymptotically adS space-time, were found some time ago \cite{VanderBij:2001ah}\footnote{Solutions of the ${\mathfrak {su}}(2)$ EYMH equations in adS with a triplet Higgs field in the adjoint representation of the gauge group have also been found \cite{Lugo:1999ai,Lugo:1999fm}.}.
The solutions resemble the asymptotically flat EYMH solitons and black holes studied in \cite{Greene:1992fw} rather than the asymptotically adS pure EYM configurations.
By a simple extension of the analysis in \cite{Boschung:1994tc}, it can be shown that the asymptotically adS EYMH solitons have an instability in the odd-parity sector \cite{VanderBij:2001ah} analogous to the instability of the corresponding asymptotically flat EYMH solitons.
Given this result, and the similarity between the asymptotically flat and adS EYMH black holes, the authors of \cite{VanderBij:2001ah} conjecture that the black hole solutions will also be unstable, but do not provide a proof since the techniques used in \cite{Winstanley:1995iq,Mavromatos:1995kc} to prove the instability of the asymptotically flat EYMH black holes do not extend to the asymptotically adS case.

In this note we close this gap by presenting a proof of the instability of black holes in ${\mathfrak {su}}(2)$ EYMH theory with a doublet Higgs field in asymptotically adS space-time under odd-parity, linear, spherically symmetric perturbations.
In section \ref{sec:eqns} we outline the equilibrium and perturbation equations satisfied by the sphaleron black holes, following \cite{Boschung:1994tc,Winstanley:1995iq,Mavromatos:1995kc}.
Our instability proof is in section \ref{sec:proof} followed by brief conclusions in section \ref{sec:conc}.

\section{Static and perturbation equations for EYMH theory in adS}
\label{sec:eqns}

We consider EYMH theory in four-dimensional, asymptotically adS space-time.
The gauge group is ${\mathfrak {su}}(2)$ and the doublet Higgs field is in the fundamental representation.
We focus on spherically symmetric soliton and black hole configurations with metric
\begin{equation}
ds^{2} = - N(t,r) S^{2}(t,r) dt^{2} + N^{-1} dr^{2} + r^{2} \left( d\theta ^{2} + \sin ^{2} \theta \, d\varphi ^{2} \right) .
\label{eq:metric}
\end{equation}
For the ${\mathfrak {su}}(2)$ gauge field, we employ the ansatz \cite{Greene:1992fw,Boschung:1994tc,VanderBij:2001ah}
\begin{align}
A  = & a_{0}(t,r)\tau _{r} dt + a_{1}(t,r) \tau _{r} dr
\nonumber \\  &
 + \left[ \omega (t,r) +1\right] \left[ -\tau _{\varphi } d\theta + \tau _{\theta } \sin \theta \, d\varphi \right]
\nonumber \\
&  + {\tilde {\omega }}(t,r) \left[ \tau _{\theta } d\theta + \tau _{\varphi } \sin \theta \, d\varphi \right] ,
\end{align}
where the $\tau _{i}$ are generators of the ${\mathfrak {su}}(2)$ gauge group in spherical coordinates (see, for example, Appendix A of \cite{Mavromatos:1995kc}).
The doublet Higgs field takes the form \cite{Greene:1992fw,Boschung:1994tc,VanderBij:2001ah}
\begin{equation}
\Phi = \frac {1}{{\sqrt {2}}} \left( \begin{array}{c}
\psi _{2} + i\psi _{1}
\\
\phi (t,r) - i\psi _{3}
\end{array} \right) ,
\end{equation}
where
\begin{equation}
{\bmath {\psi }} = \psi (t,r) {\hat {\bmath {r}}}.
\end{equation}
The Higgs potential is
\begin{equation}
V(\Phi ) = \frac {\lambda }{4} \left( \Phi ^{\dagger } \Phi - v^{2} \right) ^{2},
\end{equation}
where $\lambda $ and $v$ are constants.

Spherically symmetric EYMH solitons and black holes are described by the real quantities $N$, $S$, $a_{0}$, $a_{1}$, ${\tilde {\omega }}$, $\omega $, $\phi $ and $\psi $.
For static equilibrium configurations, all these quantities are functions of the radial coordinate $r$ only.
Furthermore, in this case we have
$a_{0}=a_{1}={\tilde {\omega }}=\psi =0$ \cite{Greene:1992fw,VanderBij:2001ah}.
The remaining nonzero matter field functions $\omega (r)$, $\phi (r)$ satisfy the following static field equations, which arise from the Yang-Mills and Higgs equations \cite{Greene:1992fw,VanderBij:2001ah}:
\begin{subequations}
\label{eq:equil}
\begin{align}
& N\omega '' + \frac {\left( NS \right)'}{S} \omega '  =  \frac {1}{r^{2}} \left( \omega ^{2}-1\right) \omega + \frac {\phi ^{2}}{4} \left( 1 + \omega \right) ,
\\
& N\phi '' + \frac {\left( NS \right) '}{S} \phi ' + \frac {2N}{r} \phi '  =  \frac {1}{2r^{2}} \phi \left( 1 + \omega \right) ^{2} + \lambda \phi \left( \phi ^{2} - v^{2} \right) ,
\end{align}
\end{subequations}
where a prime $'$ denotes differentiation with respect to $r$.
The derivatives of the metric functions $N$ and $S$ can be written in terms of the matter field functions using the Einstein equations; we shall not require these equations for our analysis.

The variational method we employ in the next section does not depend on the details of the equilibrium matter fields, but the boundary conditions they satisfy will be crucial.  We consider only the space-time exterior to a regular, nonextremal event horizon at $r=r_{h}$, in a neighbourhood of which the field variables take the form \cite{VanderBij:2001ah}
\begin{align}
& N(r) = O(r-r_{h}), \qquad S(r) = S_{h} + O(r-r_{h}),
\nonumber \\
& \omega (r) = \omega _{h} + O(r-r_{h}), \qquad
\phi (r)= \phi _{h} +O(r-r_{h}),
\label{eq:bch}
\end{align}
where $S_{h}$, $\omega _{h}$ and $\phi _{h}$ are constants.
As $r\rightarrow \infty $, the space-time metric (\ref{eq:metric}) tends to that of pure adS space-time, so that
\begin{equation}
N(r) = \frac {r^{2}}{\ell ^{2}} + 1 + O(r^{-1}), \qquad S(r) = 1 + o(r^{-1}),
\label{eq:bcinf1}
\end{equation}
where $\ell $ is the adS radius of curvature.
The matter fields have a complicated power-law decay as $r\rightarrow \infty $ \cite{VanderBij:2001ah}:
\begin{equation}
 \omega (r) = -1 + \frac {c_{1}}{r^{k_{1}}} ,  \qquad
\phi (r) = \pm v + \frac {c_{2}}{r^{k_{2}}},
\label{eq:bcinf}
\end{equation}
where $c_{1}$ and $c_{2}$ are constants, and the powers $k_{1}$ and $k_{2}$ are given by \cite{VanderBij:2001ah}
\begin{equation}
k_{1} = \frac {1}{2} \left( 1 +  {\sqrt {1+v^{2}\ell ^{2}}}\right) ,
\quad
k_{2} = \frac {3}{2} \left( 1 +  {\sqrt {1 + \frac {8\lambda v^{2}\ell ^{2} }{9}}} \right) .
\end{equation}

The boundary conditions (\ref{eq:bcinf}) at infinity constrain the YMH matter fields to have their vacuum values (as happens in the asymptotically flat case \cite{Greene:1992fw}), in contrast to the boundary conditions at infinity for pure EYM in adS, which do not constrain the value of the gauge field function $\omega $ as $r\rightarrow \infty $.
This indicates that the asymptotically adS EYMH solitons and black holes are more like their counterparts in asymptotically flat space-time than the pure EYM solutions in adS.
In asymptotically flat space-time, the matter functions $\omega $ and $\phi $ have an exponential rather than power-law fall off as $r\rightarrow \infty $ \cite{Greene:1992fw}, but this does not have a major effect on the equilibrium solutions.

We now consider linear, spherically symmetric, perturbations of the static equilibrium configurations.  By a choice of gauge, the perturbation $\delta a_{0}$ can be set to vanish identically \cite{Boschung:1994tc}.  The remaining perturbations decouple into two sectors: the even-parity (gravitational) sector consists of the perturbations of the metric functions $\delta N$ and $\delta S$, together with the matter perturbations $\delta \omega $ and $\delta \phi$; the odd-parity (sphaleronic) sector contains the perturbations $\delta a_{1}$, $\delta {\tilde {\omega }}$ and $\delta \psi $. We consider only the latter sector of perturbations. All perturbations depend on time $t$ as well as the radial coordinate $r$.

The linear perturbation equations for the odd-parity sector are the same in the asymptotically adS case as they are for the asymptotically flat case \cite{Boschung:1994tc,VanderBij:2001ah}. Defining a vector of perturbations $\Psi (t,r)$ by
\begin{equation}
\Psi (t,r) = \left( \delta a_{1}, \delta {\tilde {\omega }}, \delta \psi \right) ^{T},
\label{eq:perts}
\end{equation}
they take the form \cite{Boschung:1994tc,Mavromatos:1995kc}
\begin{equation}
- {\mathcal {A}} {\ddot {\Psi }} = {\mathcal {H}} \Psi ,
\label{eq:pert}
\end{equation}
where a dot ${\dot {}}$ denotes differentiation with respect to time $t$.
The perturbation equations (\ref{eq:pert}) involve two operators on the space of perturbations $\Psi $, namely \cite{Boschung:1994tc,Mavromatos:1995kc}
\begin{equation}
{\mathcal {A}} = \left( \begin{array}{ccc}
Nr^{2} & 0 & 0  \\
0 & 2 & 0 \\
0 & 0 & r^{2}
\end{array}
\right) ,
\label{eq:Adef}
\end{equation}
while the components of the operator ${\mathcal {H}}$ are given by \cite{Boschung:1994tc,Mavromatos:1995kc}
\begin{align}
{\mathcal {H}}_{a_{1}a_{1}} & =
2\left( N S \right) ^{2} \left( \omega ^{2} + \frac {r^{2}\phi ^{2}}{8} \right) ,
\nonumber \\
{\mathcal {H}}_{{\tilde {\omega }}{\tilde {\omega }}} & =
2p_{*}^{2} + 2NS^{2} \left( \frac {\omega ^{2}-1}{r^{2}} + \frac {\phi ^{2}}{4} \right) ,
\nonumber \\
{\mathcal {H}}_{\psi \psi } & =
2p_{*} \frac {r^{2}}{2} p_{*} + 2NS^{2} \left[
\frac {\left( 1 - \omega \right) ^{2}}{4} + \frac {r^{2}\lambda \left( \phi ^{2} -v^{2} \right) }{2} \right] ,
\nonumber \\
{\mathcal {H}}_{a_{1}{\tilde {\omega }}} & =
-2iNS \left[ \left( p_{*}\omega  \right) - \omega p_{*} \right] ,
\nonumber \\
{\mathcal {H}}_{{\tilde {\omega }}a_{1}} & =
-2i \left[ p_{*}NS\omega + NS\left( p_{*}\omega \right) \right] ,
\nonumber \\
{\mathcal {H}}_{a_{1}\psi } & =
\frac {ir^{2}NS}{2} \left[ \left( p_{*} \phi \right) - \phi p_{*} \right] ,
\nonumber \\
{\mathcal {H}}_{\psi a_{1}} & =   ip_{*} \frac {r^{2}}{2} NS\phi + \frac {ir^{2}}{2} NS \left( p_{*}\phi \right) ,
\nonumber \\
{\mathcal {H}}_{\psi {\tilde {\omega }}} & =  {\mathcal {H}}_{{\tilde {\omega }}\psi }
= -\phi NS^{2}.
\label{eq:Hdef}
\end{align}
In (\ref{eq:Adef}, \ref{eq:Hdef}), all field variables are equilibrium quantities depending on the radial coordinate $r$ only, and we have defined the differential operator
\begin{equation}
p_{*} = -iNS \frac {d}{dr}.
\end{equation}

\section{Proof of instability of sphaleron black holes in adS}
\label{sec:proof}

We will now prove that the sphaleron black hole solutions of ${\mathfrak {su}}(2)$ EYMH theory in four-dimensional asymptotically adS space-time possess an instability in the odd-parity sector of perturbations.  Our proof is a minor modification of that for the corresponding asymptotically flat black holes \cite{Winstanley:1995iq,Mavromatos:1995kc}.

We begin by considering time-periodic perturbations (\ref{eq:perts}) with frequency $\sigma $:
\begin{equation}
\Psi (t,r) = \Psi (r) e^{i\sigma t}, \qquad
\Psi (r) = \left( \delta a_{1}(r), \delta {\tilde {\omega }}(r), \delta \psi (r) \right) ^{T}.
\label{eq:timep}
\end{equation}
The perturbation equations (\ref{eq:pert}) then take the form of an eigenvalue problem for $\sigma ^{2}$:
\begin{equation}
\sigma ^{2} {\mathcal {A}} \Psi (r) = {\mathcal {H}}\Psi (r).
\label{eq:eigen}
\end{equation}
Rather than attempting to solve the above eigenvalue equation directly, we follow \cite{Volkov:1994dq,Boschung:1994tc,Winstanley:1995iq} and use a variational method to show that (\ref{eq:eigen}) possesses negative eigenvalues.
If $\sigma ^{2}<0$, the frequency $\sigma $ is purely imaginary and the perturbations (\ref{eq:timep}) grow exponentially with time, indicating an instability of the corresponding equilibrium configurations.

We consider the following inner product on the space of perturbations $\Psi (r)$:
\begin{equation}
\langle \Psi | \Upsilon \rangle = \int _{r=r_{h}}^{\infty } {\overline {\Psi }} \Upsilon \frac {dr}{NS},
\label{eq:inner}
\end{equation}
with respect to which the operators ${\mathcal {A}}$ (\ref{eq:Adef}) and ${\mathcal {H}}$ (\ref{eq:Hdef}) are symmetric (when acting on perturbations satisfying appropriate boundary conditions) and it is straightforward
to show that ${\mathcal {A}}$ is positive definite.
Our variational approach involves the functional
\begin{equation}
\sigma ^{2}(\Psi ) = \frac {\langle \Psi | {\mathcal {H}} | \Psi \rangle }{\langle \Psi | {\mathcal {A}} | \Psi \rangle }
\end{equation}
which is defined for any {\em {trial}} perturbation $\Psi (r)$ (not necessarily an eigenvector of ${\mathcal {H}}$).
The lowest eigenvalue of the operator ${\mathcal {H}}$ gives a lower bound for the functional $\sigma ^{2}(\Psi )$.
Therefore, if we can find a trial perturbation $\Psi (r)$ for which
\begin{equation}
\sigma ^{2}(\Psi ) <0, \qquad \langle \Psi | {\mathcal {A}} | \Psi \rangle < \infty ,
\label{eq:conds}
\end{equation}
then the operator ${\mathcal {H}}$ has at least one negative eigenvalue, and we have proven instability.
As emphasized in \cite{Volkov:1994dq}, the second condition in (\ref{eq:conds}) is essential for ensuring that the trial perturbations considered are normalizable, since the existence of non-normalizable perturbations $\Psi $ for which $\sigma ^{2}(\Psi )<0$ does not imply the instability of the equilibrium configurations.

Following \cite{Winstanley:1995iq,Mavromatos:1995kc}, we consider the following trial perturbations:
\begin{align}
\delta a_{1} & =  -\omega ' Z,
\nonumber \\
\delta {\tilde {\omega }} & =  \left( \omega ^{2} -1 \right) Z,
\nonumber \\
\delta \psi & =  -\frac {1}{2}\phi \left( 1 + \omega \right) Z,
\end{align}
where $Z$ is a function of $r$ to be determined shortly.
Then
\begin{equation}
\langle \Psi | {\mathcal {A}} | \Psi \rangle = \int _{r_{h}}^{\infty } \frac {Z^{2} \,  dr}{NS} \left[ Nr^{2}\omega '^{2} + 2\left( \omega ^{2}-1\right) ^{2}
+ \frac {r^{2}\phi ^{2}}{4} \left( 1 + \omega \right) ^{2} \right] ,
\label{eq:Aexp}
\end{equation}
and, using (\ref{eq:equil}) and performing an integration by parts,
\begin{multline}
\langle \Psi | {\mathcal {H}} | \Psi \rangle  =
 \\
 - \int _{r_{h}}^{\infty } S \, dr \left[ 2N\omega '^{2} + \frac {2}{r^{2}}\left( \omega ^{2}-1\right) ^{2}
+ \frac {1}{2} \phi ^{2} \left( 1 + \omega \right) ^{2}  \right]
 \\
+ \int _{r_{h}}^{\infty } S \left( 1 - Z^{2} \right) dr \left[ 2N\omega '^{2} + \frac {2}{r^{2}} \left( \omega ^{2}-1 \right) ^{2}
+ \frac {1}{2} \phi ^{2} \left( 1 + \omega \right) ^{2} \right]
 \\
+ \int _{r_{h}}^{\infty } NS \left( \frac {dZ}{dr} \right) ^{2} dr \left[ 2\left( \omega ^{2} -1 \right) ^{2} + \frac {r^{2}\phi ^{2}}{4} \left( 1 + \omega \right) ^{2} \right]
 \\
- \left[ N S Z \frac {dZ}{dr} \left\{ 2 \left( \omega ^{2}-1 \right) ^{2} + \frac {r^{2}\phi ^{2}}{4} \left( 1 + \omega \right) ^{2} \right\} \right] _{r_{h}}^{\infty } ,
\label{eq:Hexp}
\end{multline}
where we have explicitly retained the boundary terms omitted in \cite{Winstanley:1995iq,Mavromatos:1995kc}.

For equilibrium solitons, the lower limit on the integral in (\ref{eq:inner}) is set to be $r=0$ rather than $r=r_{h}$.   In this case it is sufficient to simply set $Z\equiv 1$ \cite{VanderBij:2001ah}; the boundary conditions at infinity (\ref{eq:bcinf1}, \ref{eq:bcinf}) and at the origin ensure the finiteness of $\langle \Psi | {\mathcal {A}} | \Psi \rangle $ (\ref{eq:Aexp}), and in $\langle \Psi | {\mathcal {H}} | \Psi \rangle $ all terms except the first integral (which is manifestly negative) vanish, so that $\langle \Psi | {\mathcal {H}} | \Psi \rangle <0$ and instability is proven.

Setting $Z\equiv 1$ does not work for the black hole case because then the integrand in (\ref{eq:Aexp}) would diverge in a nonintegrable way as $r\rightarrow r_{h}$.  We therefore need to define a suitable function $Z$.  First define the usual ``tortoise'' coordinate $r_{*}$ by
\begin{equation}
\frac {dr_{*}}{dr} = \frac {1}{NS}.
\end{equation}
As $r\rightarrow r_{h}$, the tortoise coordinate $r_{*}\rightarrow -\infty $.  However, as $r\rightarrow \infty $, the boundary conditions (\ref{eq:bcinf1}) mean that $r_{*}$ tends to a constant, which may be taken to be zero without loss of generality.
The functions $Z$ used in \cite{Volkov:1994dq,Winstanley:1995iq,Mavromatos:1995kc} to prove instability for sphaleron black holes in the asymptotically flat case assume that $r_{*}$ has values in the full range $(-\infty , \infty )$ and so cannot be used here.
However, a  minor modification is all that is required.

To this end, we define a sequence of functions $Z_{k}$ in terms of $r_{*}$, as follows (cf.~\cite{Volkov:1994dq,Winstanley:1995iq,Mavromatos:1995kc}):
\begin{equation}
Z_{k}(r_{*}) = Z\left( \frac {r_{*}}{k} \right), \qquad k = 1,2, \ldots ,
\end{equation}
where $Z(r_{*})$ is defined by
\begin{equation}
Z(r_{*}) = 1 \quad {\mbox {for $r_{*} \in [-a, 0]$}}, \qquad
Z(r_{*}) = 0 \quad {\mbox {for $r_{*} < -a-1$}} ,
\end{equation}
for some positive constant $a>0$, and furthermore there is another positive constant $D>0$ such that
\begin{equation}
0 \le \frac {dZ}{dr_{*}} \le D \quad {\mbox {for $r_{*} \in [-a-1, -a] $}}.
\end{equation}
As $r\rightarrow \infty $, for each $k$, we have $Z_{k}=1$ and $\frac {dZ_{k}}{dr}=0$ so that the contribution to the boundary term in (\ref{eq:Hexp}) coming from infinity vanishes. These facts and the boundary conditions (\ref{eq:bcinf1}, \ref{eq:bcinf}) ensure that the integrands in (\ref{eq:Aexp}, \ref{eq:Hexp}) all tend to zero as $r\rightarrow \infty $, and yield finite integrals.
As $r\rightarrow r_{h}$, for each $k$ it is the case that $Z_{k}=0$ and $\frac {dZ_{k}}{dr} =0$.
These, together with the boundary conditions (\ref{eq:bch}), ensure that all integrals in (\ref{eq:Aexp}, \ref{eq:Hexp}) are finite and that the contribution to the boundary term in (\ref{eq:Hexp}) from the horizon also vanishes.
In particular, for each $Z_{k}$, we have that $\langle \Psi | {\mathcal {A}} | \Psi \rangle < \infty $, as required.

The first integral in $\langle \Psi | {\mathcal {H}} | \Psi \rangle $ (\ref{eq:Hdef}) is clearly negative.  Write the second and third as follows:
\begin{equation}
I_{2} = \int _{r_{h}}^{\infty } dr \left( 1 - Z_{k}^{2} \right) {\mathcal {F}}, \qquad
I_{3} = \int _{r_{h}}^{\infty } NS \, dr \left( \frac {dZ_{k}}{dr} \right) ^{2} {\mathcal {G}}  ,
\end{equation}
where the positive functions ${\mathcal {F}}$ and ${\mathcal {G}}$ are given by
\begin{align}
{\mathcal {F}} & =  S \left[ 2N\omega '^{2}+ \frac {2}{r^{2}}\left( \omega ^{2}-1 \right) ^{2} + \frac {\phi ^{2}}{2} \left( \omega - 1 \right) ^{2} \right] ,
\nonumber \\
{\mathcal {G}} & =  2\left( \omega ^{2}-1\right) ^{2} + \frac {1}{4} r^{2} \phi ^{2} \left( \omega -1 \right) ^{2} .
\end{align}
It is straightforward to show that
\begin{equation}
0 \le I_{2} \le \left( r_{k}-r_{h} \right) {\mathcal {F}}_{M}, \qquad
0 \le I_{3} \le \frac {D^{2}}{k} {\mathcal {G}}_{M} ,
\label{eq:Ibounds}
\end{equation}
where $r_{*}(r_{k})=-ka$ and
\begin{equation}
{\mathcal {F}}_{M} = \max _{r\in [r_{h},\infty )} {\mathcal {F}} ,
\qquad
{\mathcal {G}}_{M} = \max _{r\in [r_{h},\infty )} {\mathcal {G}} .
\end{equation}
The boundary conditions (\ref{eq:bch}, \ref{eq:bcinf1}, \ref{eq:bcinf}) ensure the finiteness of ${\mathcal {F}}_{M}$ and ${\mathcal {G}}_{M}$.
The bounds on the right-hand-side of each inequality in (\ref{eq:Ibounds}) can be made arbitrarily small by considering sufficiently large $k$: for the integral $I_{2}$ this is because $r_{k}\rightarrow r_{h}$ as $k\rightarrow \infty $.
Therefore, for sufficiently large $k$ the dominant contribution to $\langle \Psi | {\mathcal {H}} | \Psi \rangle $ (\ref{eq:Hdef}) comes from the first integral and $\langle \Psi | {\mathcal {H}} | \Psi \rangle <0$.
This suffices to prove instability.

\section{Conclusions}
\label{sec:conc}

In this paper we have proven that static, spherically symmetric, sphaleron black holes in ${\mathfrak {su}}(2)$ EYMH theory with a doublet Higgs field in the fundamental representation, in four-dimensional asymptotically adS space-time, are unstable.
Coupled with the analysis in \cite{VanderBij:2001ah}, we conclude that both solitons and black holes in this theory in adS are unstable, like their asymptotically flat counterparts.
This is in contrast to the situation in pure ${\mathfrak {su}}(2)$ EYM theory in adS, where there exist stable solitons and black holes \cite{Winstanley:1998sn,Bjoraker:1999yd,Bjoraker:2000qd}.
It is also interesting to note that there are stable black hole solutions of pure Einstein-Higgs theory (with no gauge field) in four-dimensional, asymptotically adS space-time \cite{Torii:2001pg}.

How can we understand this difference in behaviour?
Mathematically, the key difference between the pure EYM theory and EYMH theory in adS is the boundary conditions on the gauge field at infinity, the boundary conditions for EYMH being much more restrictive (fixing the value of $\omega $ as $r\rightarrow \infty $) than in the EYM case (where $\omega $ can take any finite value as $r\rightarrow \infty $).
Physically, in EYMH theory the gauge field dynamically acquires a mass and both it and the Higgs field must be in the vacuum configuration at infinity.
In pure EYM theory, where the gauge field is massless, for stable solutions it generically is not in the vacuum configuration at infinity.
Interestingly, for stable solutions in Einstein-Higgs theory in adS, the boundary conditions on the scalar field at infinity are also very restrictive: for stable configurations the scalar field must approach the local maximum of the Higgs potential \cite{Torii:2001pg}. However, this means that the scalar field is not in the vacuum configuration at infinity.

We therefore conjecture that the boundary conditions at infinity are of importance in determining whether a particular matter model has stable soliton and hairy black hole solutions in asymptotically adS space-time. Based on the above discussion, matter fields which have to be in the vacuum configuration at infinity seem to yield only unstable solitons and hairy black holes, while those that can have nonvacuum values at infinity seem to have at least some stable equilibrium solutions.

It would be interesting to test this conjecture with other matter models in adS.  As a starting point, in a forthcoming work we will examine soliton and black hole solutions of Einstein-non-Abelian-Proca (ENAP) theory in adS \cite{Supakchai}.
In asymptotically flat space-time, solitons and black holes in ENAP theory (in which the gauge field is given an effective mass by hand in the action, rather than mass being dynamically generated by the Higgs field) share many properties with those in EYMH theory \cite{Greene:1992fw}. Like the authors of \cite{VanderBij:2001ah}, we conjecture that the same is true in asymptotically adS space-time, and will investigate this elsewhere \cite{Supakchai}.

\section*{Acknowledgments}
E.W.~thanks Supakchai Ponglertsakul for helpful discussions.
This work is supported by the Lancaster-Manchester-Sheffield Consortium for
Fundamental Physics under STFC grant ST/L000520/1.
E.W.~thanks the University of Canterbury for a Visiting Erskine Fellowship supporting this work.


\end{document}